\documentclass[twoside]{ilcws08}
\usepackage[latin1]{inputenc}
\usepackage[dvips]{graphicx,epsfig,color}
\usepackage{wrapfig,rotating}
\usepackage{amssymb,amsmath,array}

\newcommand{\hbb}{$H \to b \bar{b}$}
\newcommand{\hcc}{$H \to c \bar{c}$}
\newcommand{\eezh}{$e^{+} e^{-} \to ZH$}
\newcommand{\zhnnh}{$ZH \to \nu \bar{\nu} H$}
\newcommand{\zhnnbb}{$ZH \to \nu \bar{\nu} b \bar{b}$}
\newcommand{\zhnncc}{$ZH \to \nu \bar{\nu} c \bar{c}$}
\newcommand{\nnqq}{$\nu \nu qq$}
\newcommand{\nlqq}{$\nu \ell qq$}
\newcommand{\llqq}{$\ell \ell qq$}
\newcommand{\nnll}{$\nu \nu \ell \ell$}
\newcommand{\llll}{$\ell \ell \ell \ell$}
\begin{document}
\title{Measurement of Higgs Branching Ratio at ILC}
\author{Kohei Yoshida
\vspace{.3cm}\\
Department of Physics, Tohoku University, Sendai, Japan
}

\maketitle

\begin{abstract}
Measurement of Higgs branching ratio is necessary to investigate Higgs coupling to particle masses. Especially, it is the most important program to measure the branching ratio of \hbb~and \hcc~at the international linear collider (ILC). We have studied the measurement accuracy of Higgs branching ratio at ILC with $\sqrt{s} = 250$ GeV by using \zhnnh~events. We obtained the Higgs branching ratio with 1.1\% and 13.7\% accuracy for \hbb~and \hcc, respectively.
\end{abstract}

\section{Introduction}
In the Higgs mechanism, Higgs coupling is proportional to a particle mass. For that reason, it is important to measure the Higgs coupling to particle masses, i.e. Higgs branching ratio, is important to confirm Higgs mechanism and distinguish the Standard Model extensions. Especially, it is the most important program to measure the branching ratio of \hbb~and \hcc~at ILC \cite{rdr} with the excellent performance of the flavor tagging.

We have studied the measurement accuracy of Higgs branching ratio at ILC by using \zhnnh~events. In this paper, we report the measurement accuracy of Higgs branching ratio of \hbb~and \hcc.

\section{Simulation tools}
In this study, we used common generator samples in the ILC community for $ZH$ events and standard model backgrounds, which were prepared with WITHERD at SLAC \cite{ilcdata}. In this study, the Higgs mass was assumed to be 120 GeV. We used the center of mass energy of $\sqrt{s} = 250$ GeV and the integrated luminosity of 250 fb$^{-1}$. Here, the beam energy spread was assumed as 0.3\% for the electron and positron beam. The beam polarization was set to 80\% left-handed for the electron beam and 30\% right-handed for the positron beam. 

The signal and background events were simulated by the full simulator, Mokka\cite{mokka}, where the detector model is \verb|ILD_00| was implemented as the detector model \cite{ild}. Hadronization was done by Pythia6.409, in which the Higgs branching ratio is defined as shown in Table \ref{tb:hbr} for the Higgs mass of 120 GeV. After the detector simulation, the reconstruction was performed by Marlin\cite{marlin}. 

\begin{table}
\begin{center}
\begin{tabular}{|l|r|} \hline
& Branching ratio \\ \hline
$b\bar{b}$          & 65.7\% \\ \hline
$W^{+} W^{-}$       & 15.0\% \\ \hline
$\tau^{+} \tau^{-}$ & 8.0\% \\ \hline
$gg$                & 5.5\% \\ \hline
$c\bar{c}$          & 3.6\% \\ \hline
\end{tabular}
\caption{The Higgs branching ratio defined in Pythia6.409.}
\end{center}
\label{tb:hbr}
\end{table}

\section{Event selection}
\begin{figure}[tb]
\begin{center}
\includegraphics[width=6cm]{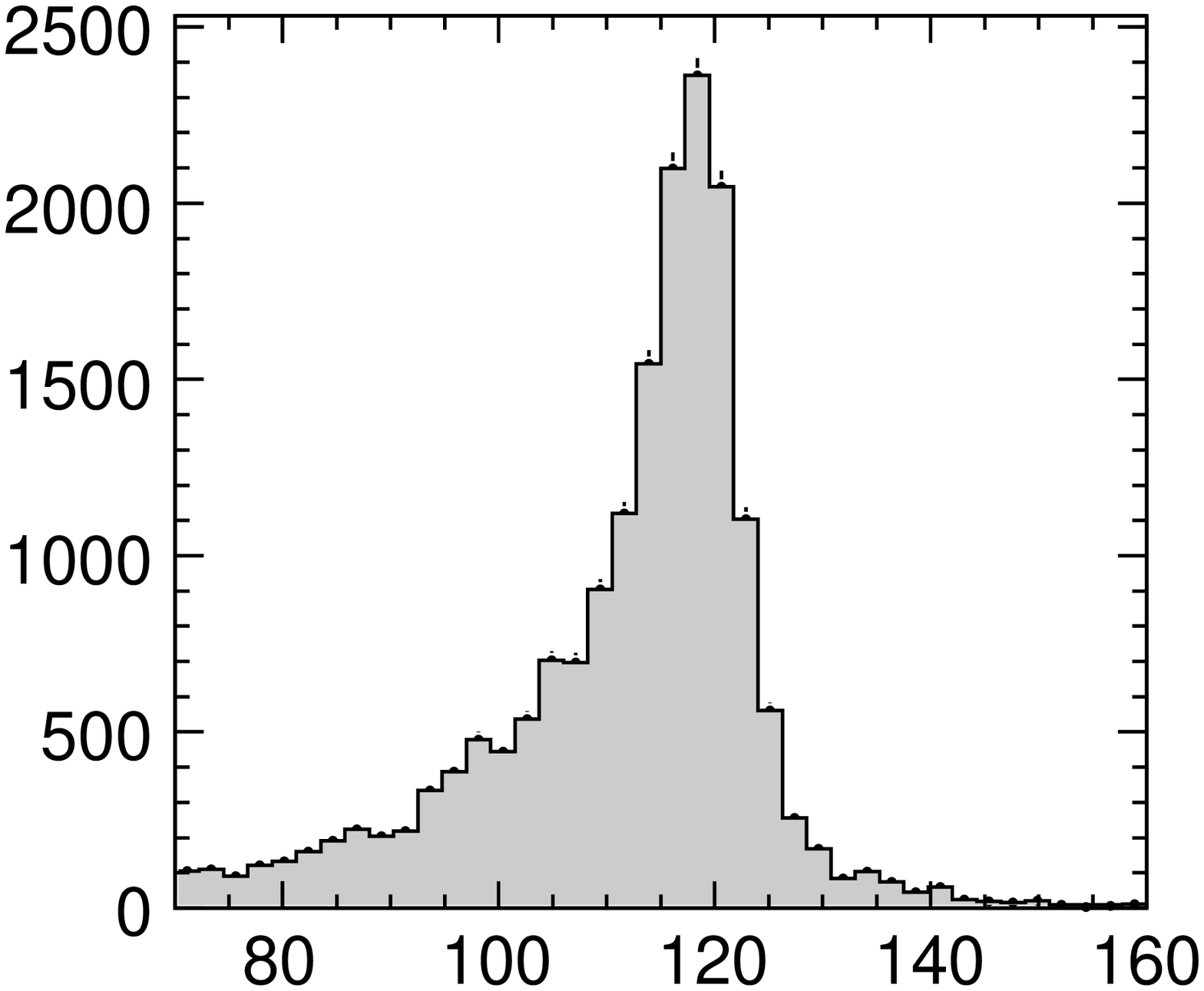}
\includegraphics[width=6cm]{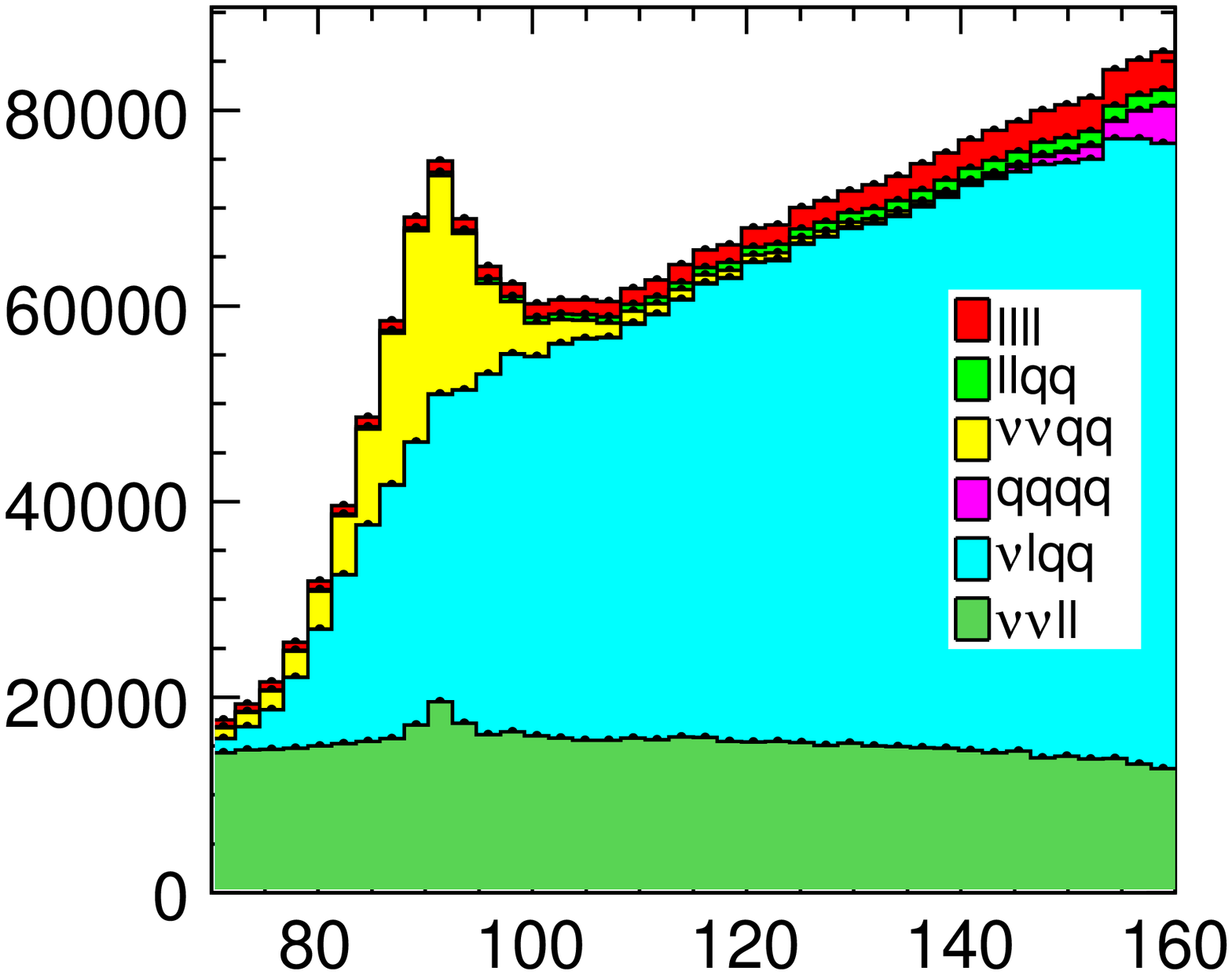}
\end{center}
\caption{Distribution of the reconstructed di-jet mass for signal (left) and background (right).}
\label{fig:hmass}
\end{figure}

In this study, the final states of four fermions are considered as background events, where they are classified into 6 groups, \nnqq, $qqqq$, \nlqq, \llqq, \nnll~and \llll.
The signal and background events are summarized in Table \ref{tb:reduction}. All events are reconstructed as 2-jet events by Durham jet algorithm \cite{durham}. By using the reconstructed 2 jets, the di-jet mass ($M_{jj}$) was reconstructed as shown in Fig.  \ref{fig:hmass}.  Since the background events dominate in the Higgs mass region, the selection cuts were investigated.

At first, we studied the distribution of missing mass ($M_{\mathrm{miss}}$). Since a Z boson decays into the neutrino pair in \zhnnh~events, the missing mass should be consistent with Z boson mass (91.2 GeV). We, therefore, selected the events with $80 \mathrm{~GeV} < M_{\mathrm{miss}} < 140 \mathrm{~GeV}$. Applying this cut, \llll, \llqq, and $qqqq$ events were suppressed. Then, we required that the reconstructed di-jet particles have the transverse momentum ($p_{\mathrm{T}}$) from 20 to 70 GeV and longitudinal momentum ($p_{\mathrm{L}}$) below 60 GeV. We selected the number of charged tracks ($N_{\mathrm{tracks}}$) above 10 to remove $W^{+}W^{-} \to l^{+} \nu l^{-} \bar{\nu}$ events.

After the selection cuts so far, ${\tau}{\nu_{\tau}}qq$ events become the main background. The maximum track momentum in each events ($p_{\mathrm{max}}$) were investigated since the charged tracks from $\tau$ have relatively higher momentum than those from $b$-jets. We selected the events with $p_{\mathrm{max}} < 30 \mathrm{~GeV}$. $Y_{+}$ is the threshold $y$-value to reconstruct 2-jet as 3-jets. Since the final state of $ZH \to \nu \bar{\nu} q \bar{q}$ and ${\tau}{\nu_{\tau}}qq$ is 2 and 3 bodies, respectively, $Y_{+}$ for $ZH \to \nu \bar{\nu} q \bar{q}$ events has smaller value than ${\tau}{\nu_{\tau}}qq$ events. On the other hand, $Y_{-}$, the $y$-value to reconstruct 2-jet as 1-jets, has larger value for $ZH \to \nu \bar{\nu} q \bar{q}$ events than $\nu \bar{\nu} qq$ and $l \nu qq$ because $\beta$ of $W$ and $Z$ bosons from decay of $WW$ and $ZZ$ events is larger than Higgs from $ZH \to \nu \bar{\nu} q \bar{q}$. We, therefore, selected $Y_{+} < 0.02$ and $0.2 < Y_{-} < 0.8$. 

\begin{figure}[tb]
\begin{center}
\includegraphics[width=6cm]{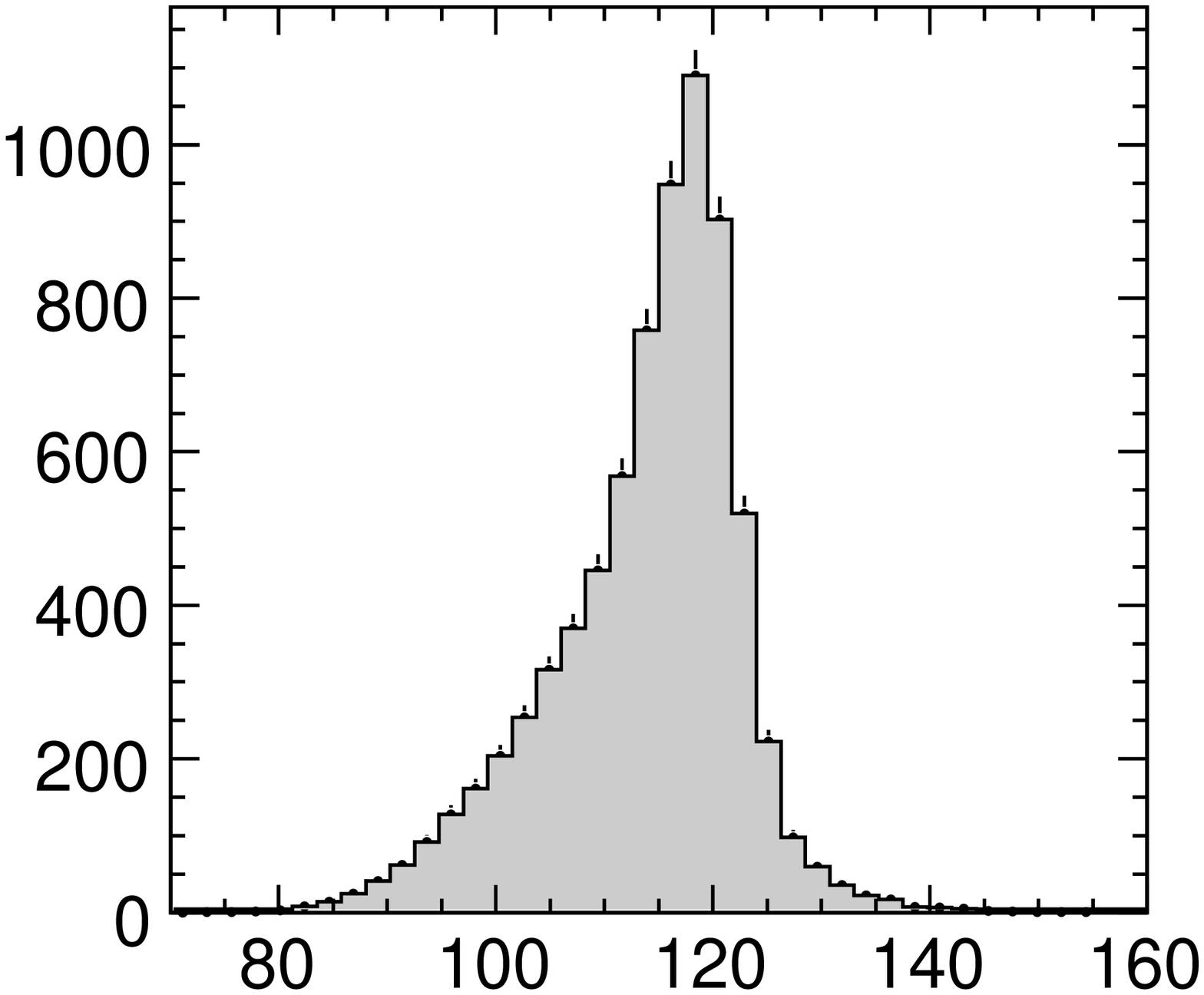}
\includegraphics[width=6cm]{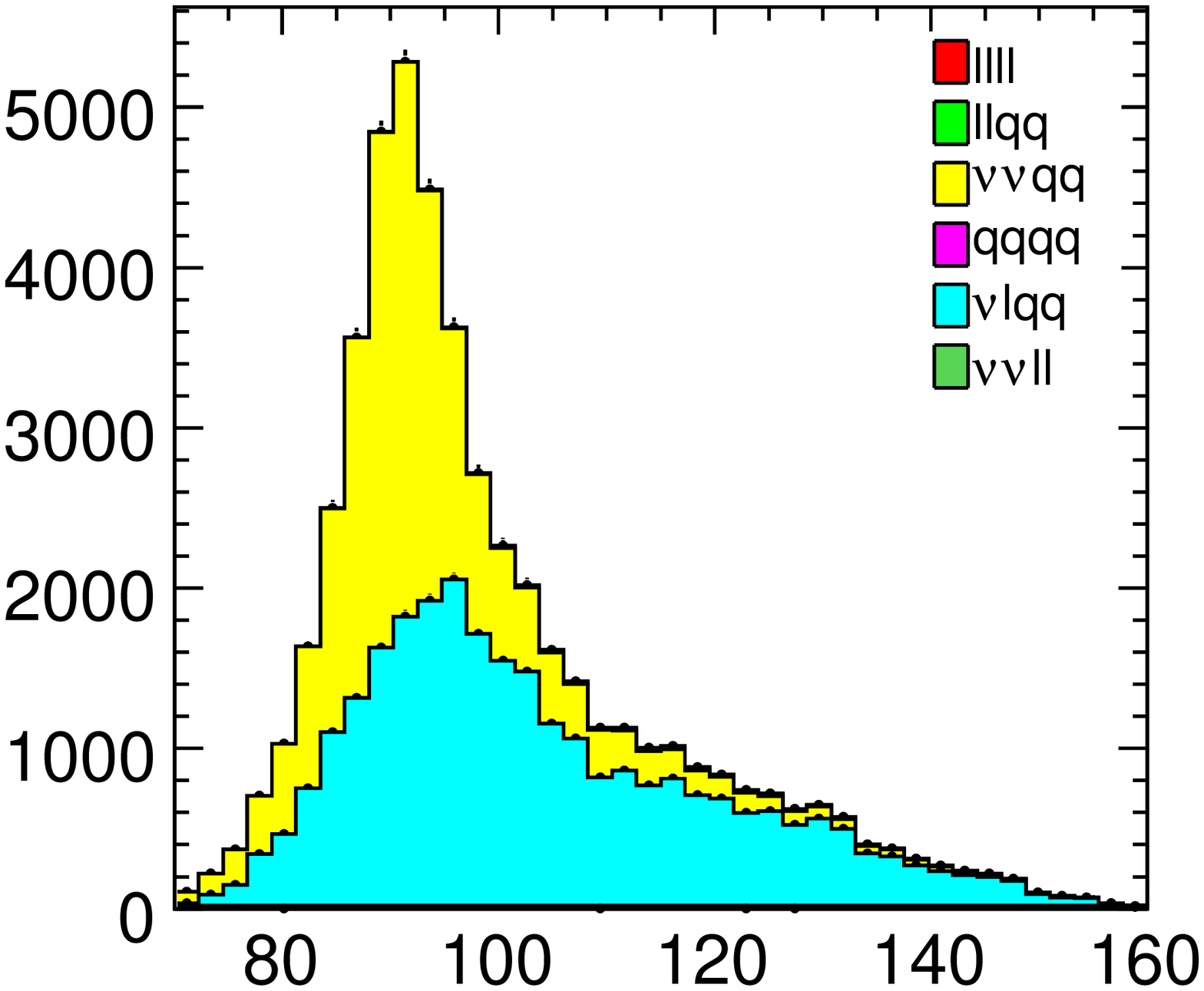}
\end{center}
\caption{Distribution of the reconstructed di-jet mass after the selection cuts for signal (left) and background (right).}
\label{fig:dijet}
\end{figure}

Finally, the signal region was set to be $100 \mathrm{~GeV} < M_{jj} < 130 \mathrm{~GeV}$. After all the selection cuts, \nnqq~events from $WW$ and $ZZ$ events were reduced as shown in Fig. \ref{fig:dijet}. The number of signal and background events and the selection efficiencies after the selection cut was summarized in Table \ref{tb:reduction}.

\begin{table}[tb]
\footnotesize
\begin{center}
\begin{tabular}{|r|r|r|r|r|}
\multicolumn{5}{l}{} \\ \hline
& cross section (fb) & No. of events & No. of events after all cuts & Efficiency (\%) \\ \hline
\zhnnh & 77.4 & 19,360 & 7,384 & 38.14 \\ \hline
\zhnnbb & 52.2 & 13,062 & 6,434 & 49.26 \\ \hline
\zhnncc & 2.83 & 707 & 318 & 44.98 \\ \hline
$\nu_eeqq$ & 5843.2 & 1,460,797 & 851 & 0.06 \\ \hline
$\nu_\mu\mu q$ & 5309.3 & 1,327,332 & 2,288 & 0.17 \\ \hline
$\nu_\tau\tau q$ & 5304.2 & 1,326,061 & 24,979 & 1.88 \\ \hline
$\nu_\nu qq$ & 599.9 & 149,979 & 21,653 & 14.44 \\ \hline
Other & 25291 & 6,322,758 & 335 & 0.01 \\ \hline
\end{tabular}
\caption{The number of events for signal and background, and the selection efficiencies after the selection cuts.}\label{tb:reduction}
\end{center}
\end{table}

\section{Measurement of Higgs branching ratio}
To measure the Higgs branching ratio of \hbb~and \hcc, the template fitting was performed \cite{temp}. For the template fitting, 3-dimensional histogram for the $b$-, $c$-, and $bc$-likeness was used, which are obtained as output values from LCFIVertex package \cite{lcfi}. In LCFIVertex, neural-net training was done by using $Z \to qq$ events at $Z$-pole (91.2 GeV) to derive $b$- and $c$-likeness. $bc$-likeness is $c$-likeness whose neural-net training is done by using only $Z \to bb$ events as background. The each flavor-likeness for two jets are combined as,
\begin{eqnarray}
X\mathrm{-likeness} = \frac{X_1{\cdot}X_2}{X_1{\cdot}X_2+(1-X_1)(1-X_2)}
\end{eqnarray}
where $X = b$ , $c$ or $bc$. $X_1$ and $X_2$ are the flavor-likeness of the first and second jet, respectively.

\begin{figure}[tb]
\begin{center}
\includegraphics[width=6cm]{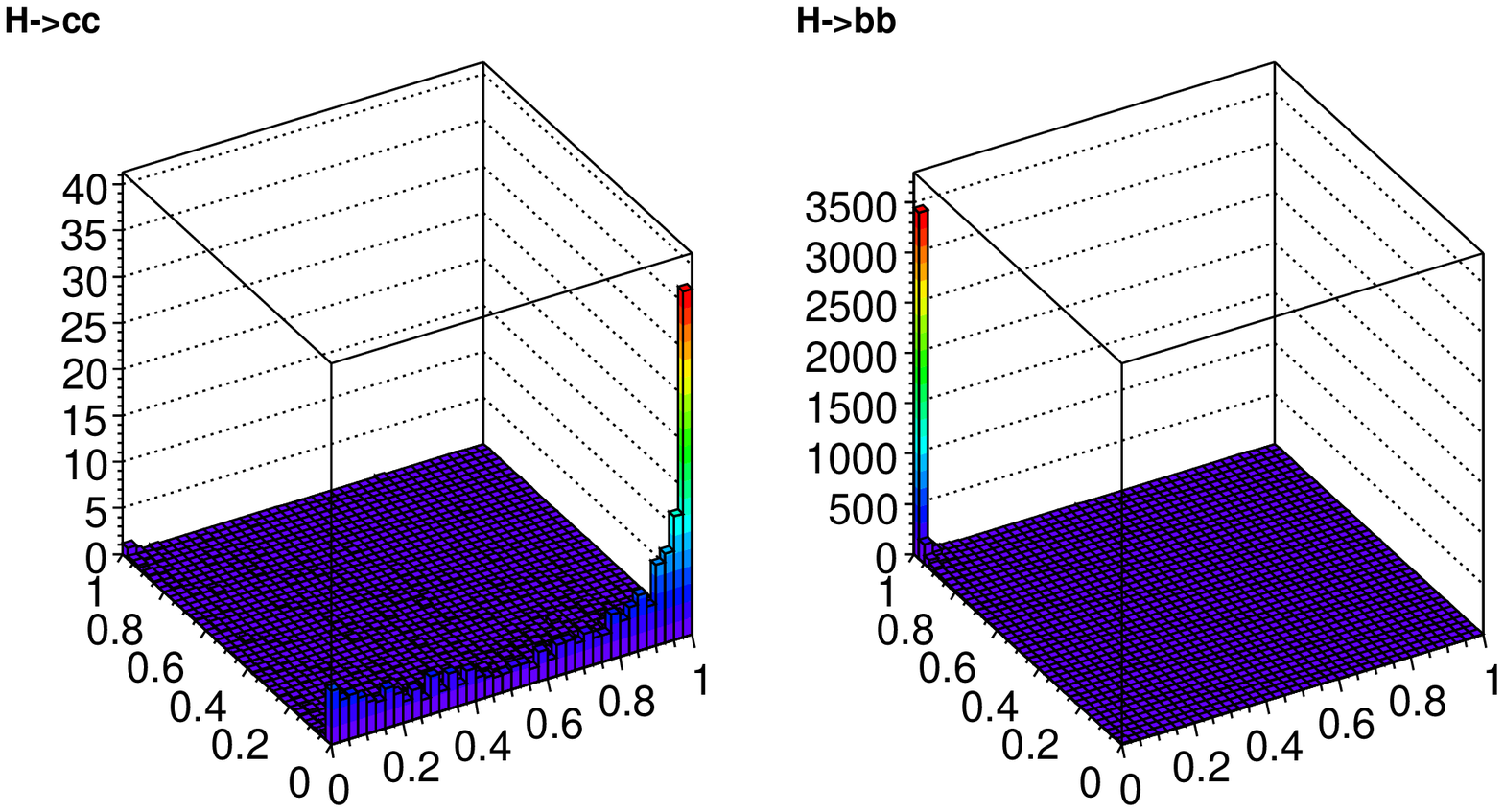}
\includegraphics[width=9cm,height=3.8cm]{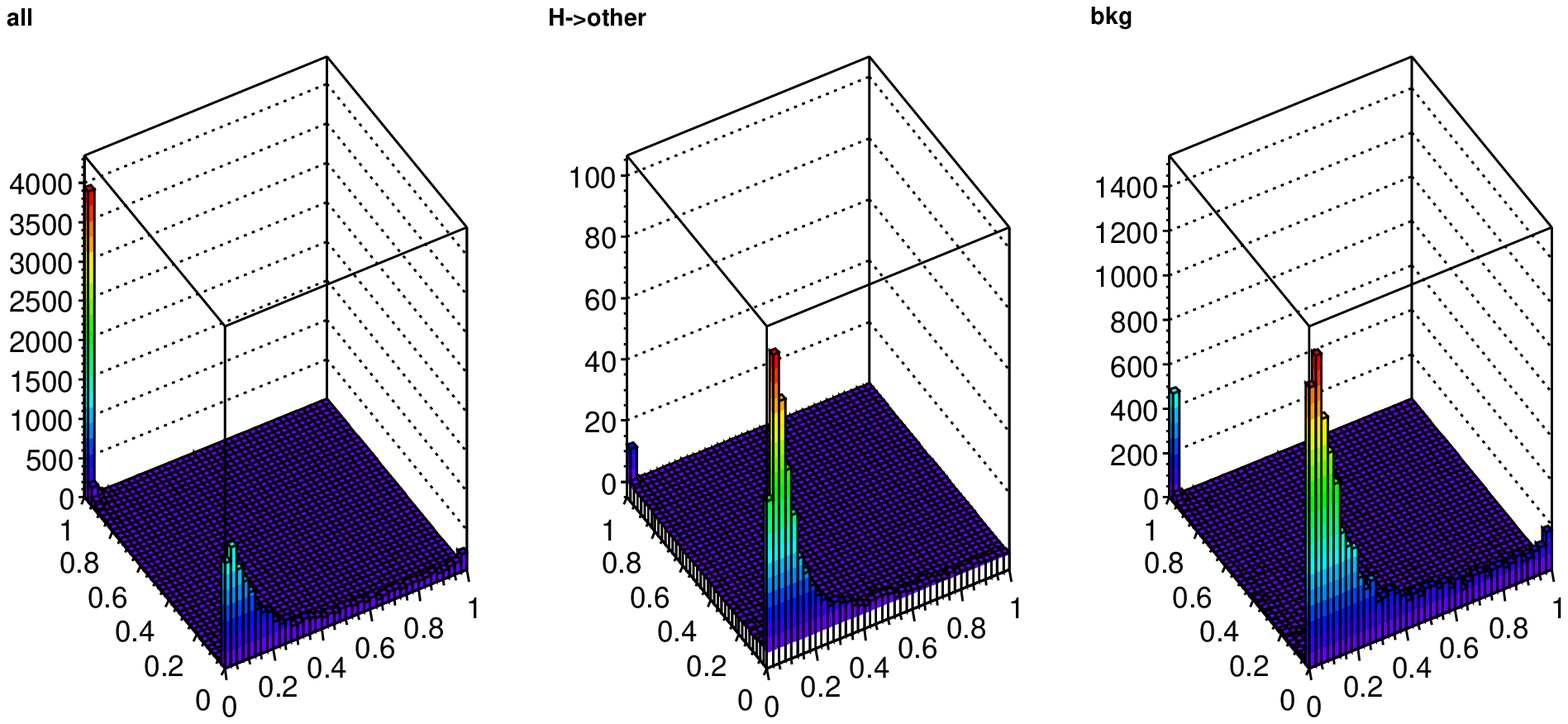}
\end{center}
\caption{2-dimensional template histogram for $b$-likeness and $c$-likeness.}
\label{fig:template}
\end{figure}

The template sample is separated into \hbb, \hcc, $H \to other$, and Standard Model background events. Figure \ref{fig:template} shows the 2-dimensional template histogram for $b$-likeness and $c$-likeness. In $H \to other$ sample, $H \to gg$ and $H \to W^-W^+$ events are dominant. Since the both distributions are identical, they are treated in one template sample.

In the template fitting, the fitting parameters ($r_{bb}$, $r_{cc}$, $r_{oth}$, and $r_{bkg}$) were adjusted to minimize the following $\chi^2$ function:
\begin{eqnarray}
\chi^2 = \sum^{n_b}_{i=1}\sum^{n_c}_{j=1}\sum^{n_{bc}}_{k=1}\frac{(N^{data}_{ijk}-\sum_sr_s(\frac{N^{ZH}}{N^s})N^s_{ijk}-r_{bkg}N^{bkg}_{ijk})^2}{N^{all}_{ijk}},
\end{eqnarray} 
where $s$ shows $b\bar{b}$, $c\bar{c}$ and $other$. $r_{bb}$, $r_{cc}$, $r_{oth}$ are the fraction of \hbb, \hcc, $H \to others$ in $ZH$ events after the selection cut, where we set $r_{other} = 1 - r_{cc} - r_{bb}$.  $r_{bkg}$ is the normalization factor of the Standard Model background. $N^s_{ijk}$ are the number of expected events in ($i$, $j$, $k$) bin of the 3-dimensional histogram.

\begin{wrapfigure}{r}{0.4\columnwidth}
\centerline{\includegraphics[width=0.37\columnwidth]{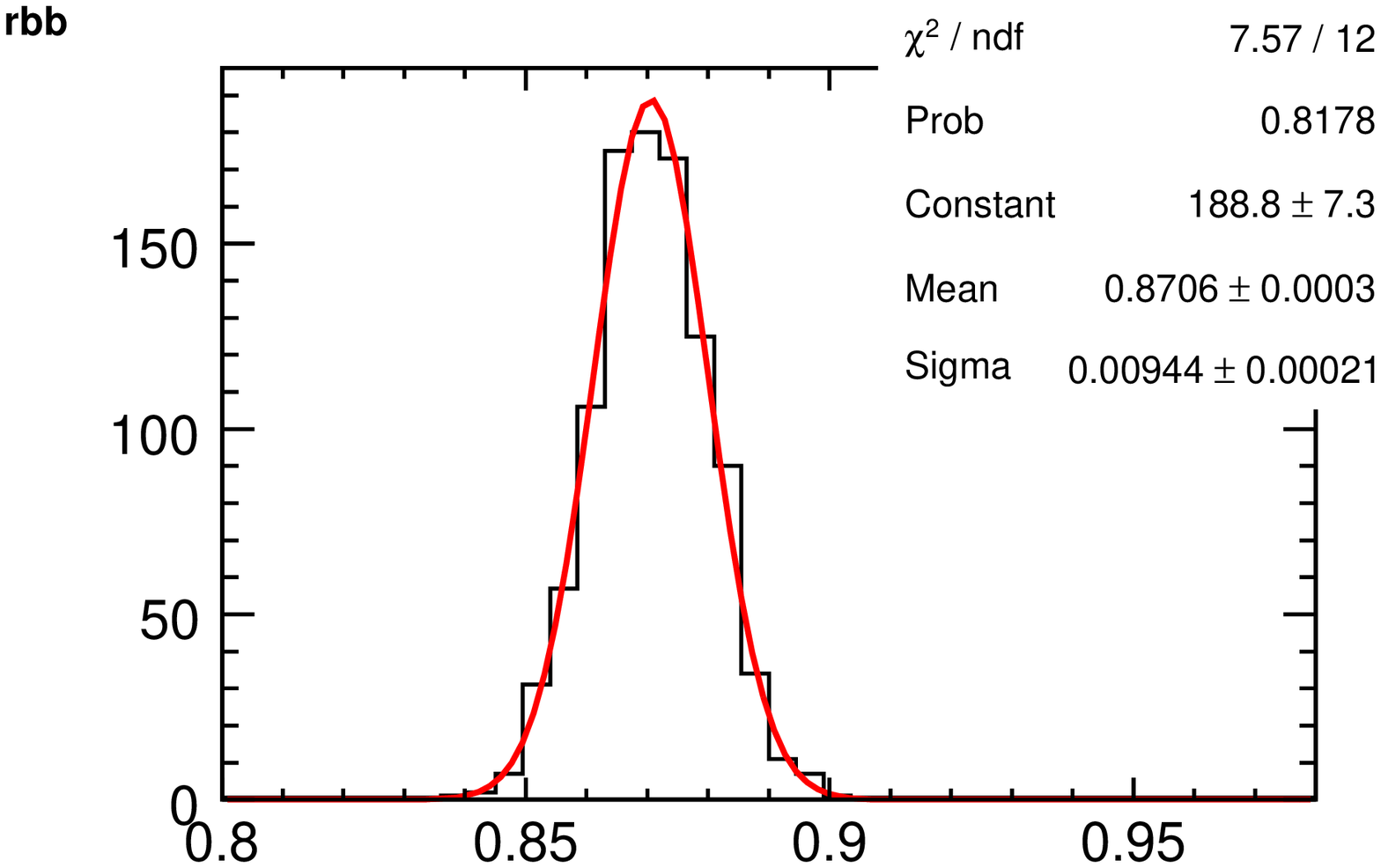}}
\centerline{\includegraphics[width=0.37\columnwidth]{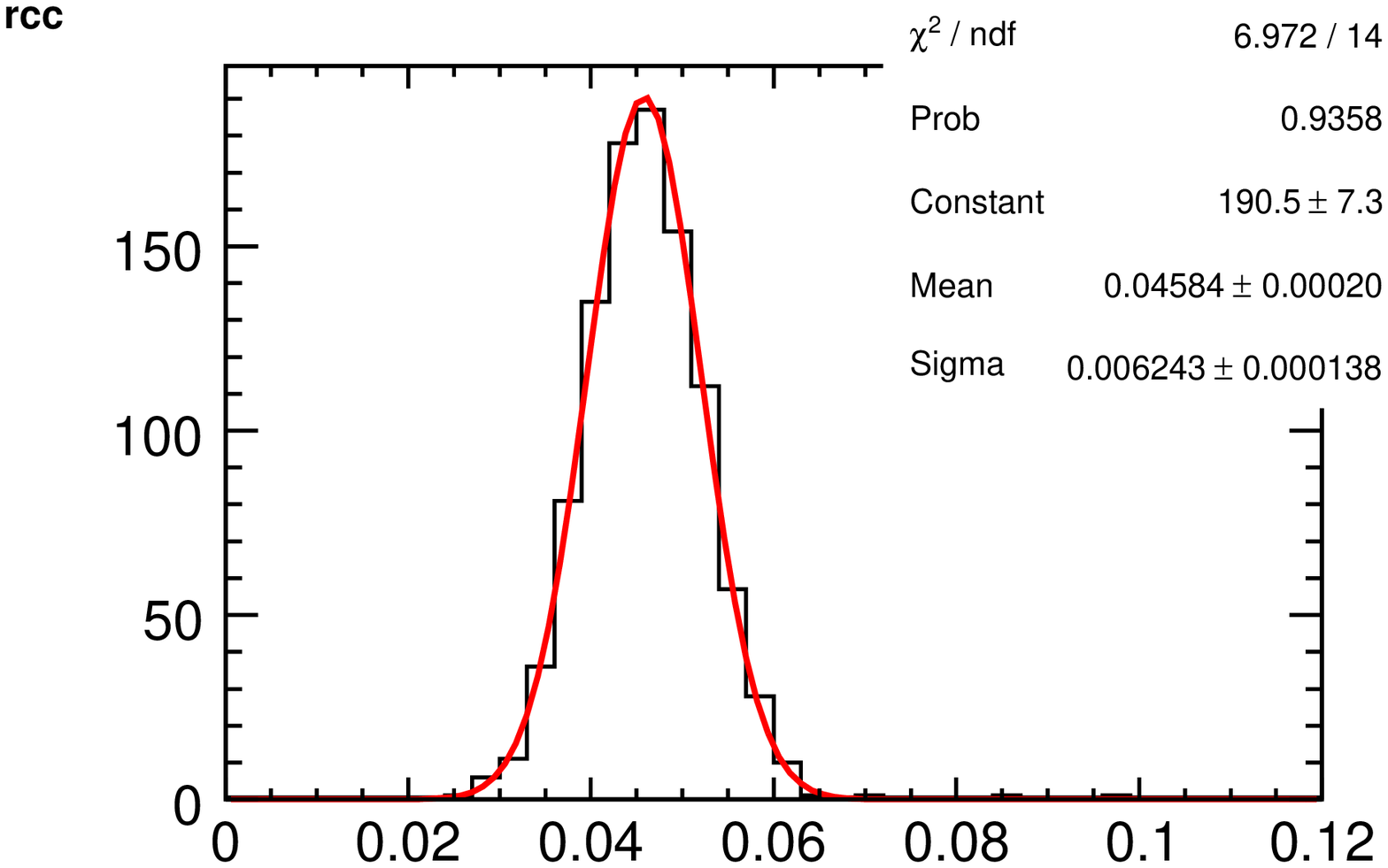}}
\caption{Distribution of $r_{bb}$ (Upper) and $r_{cc}$ (Lower) obtained by the template fitting.}
\label{fig:ratio}
\end{wrapfigure}

To estimate the reconstruction accuracy of $r_{bb}$ and $r_{cc}$, the fitting was done for 1,000 times by using Toy-MC. Figure \ref{fig:ratio} shows the distributions of $r_{bb}$ and $r_{cc}$ obtained by the fitting. $r_{bb}$ and $r_{cc}$ were determined to be $0.87 \pm 0.01$ and $0.046 \pm 0.009$, respectively. These mean values are consistent with the true $r_{bb}$ (0.87) and $r_{cc}$ (0.046). From the result, if the cross section of \eezh~can be determined with other measurements like a measurement of the Higgs recoil mass \cite{recoil} and the selection efficiencies of \zhnnbb~and \zhnncc~are known, Higgs branching ratio of \hbb~and \hcc~can be measured with accuracy of 1.1\%  and 13.7\%, respectively. 

To evaluate the influence of Standard Model background on determination of the Higgs branching ratio, we performed the template fitting, fixing $r_{bkg}$ to 1. $r_{bb}$ and $r_{cc}$ were determined to be $0.87 \pm 0.01$ and $0.046 \pm 0.006$, respectively. It corresponds to the measurement accuracy of 1.1\% and 13.6\% for $r_{bb}$ and $r_{cc}$, respectively. From this result, it was found that the fluctuation of the background normalization has only negligible effects on the measurement of Higgs branching ratio. 

Without any other measurement, we can measure the relative branching ratio between \hbb~and \hcc~by analysis of only \zhnnh~events as follows:
\begin{eqnarray}
  \frac{BR(H \to c\bar{c})}{BR(H \to b\bar{b})} = \frac{r_{cc}/\epsilon_{cc}}{r_{bb}/\epsilon_{bb}},
\end{eqnarray}
where $\epsilon_{bb}$ and $\epsilon_{cc}$ are the selection efficiencies of \hbb~and \hcc~events as shown in Table \ref{tb:reduction}. The relative branching ratio of $0.054 \pm 0.007$ was obtained for the template fitting with free and fixed $r_{bkg}$, which corresponds to 13.3\% accuracy. The measurement accuracy for Higgs branching ratio is summarized in Table \ref{tb:result}.

\begin{table}[tb]
\footnotesize
\begin{center}
\begin{tabular}{|l|r|r|} \hline
& $r_{bkg}$: free & $r_{bkg} = 1$ \\ \hline
BR(\hbb)      & 1.1\%  & 1.1\%    \\ \hline
BR(\hcc)      & 13.7\% & 13.6\%   \\ \hline
BR(\hcc/\hbb) & 13.3\% & 13.3\%   \\ \hline
\end{tabular}
\caption{The measurement accuracy of Higgs branching ratio. 
For measurement accuracy of BR(\hbb) and BR(\hcc), it is assumed that the cross section of $ZH$ is determined by other measurements.}
\label{tb:result}
\end{center}
\end{table}
\section{Conclusion}

Measurement of Higgs branching ratio is necessary to investigate Higgs coupling to particle masses. Especially, it is the most important program to measure the branching ratio of \hbb~and \hcc~at ILC. We have studied the measurement accuracy of Higgs branching ratio at ILC with $\sqrt{s} = 250$ GeV by using \zhnnh~events. For Higgs mass of 120 GeV and the integrated luminosity of 250 fb$^{-1}$, we obtained the measurement accuracy of 1.1\% and 13.7\% for \hbb~and \hcc, respectively, assuming that the cross section of $ZH$ is determined by other measurements. Finally, the relative branching ratio between \hbb~and \hcc~was obtained with 13.3\% accuracy.

\section{Acknowledgments}
The authors would like to thank all the members of the ILC physics subgroup
\cite{softg} for useful discussions and ILD optimization working group. 
This study is supported in part by the Creative Scientific Research Grant
No. 18GS0202 of the Japan Society for Promotion of Science.


\begin{footnotesize}



%

\end{footnotesize}

\end{document}